\documentstyle[12pt]{article}
\input epsf

\def\bild#1#2{    
        \vspace*{-5mm}
        \begin{center}
        \begin{math}
        \epsfxsize#2cm
        \epsffile{#1}
        \end{math}
        \end{center}
        }

\begin{document}
\renewcommand{\thefootnote}{\fnsymbol{footnote}}
%%%%%%%%%%%%%%%%%%%%%% title page %%%%%%%%%%%%%%%%%%%%%%%%%%%%%%
\begin{titlepage}
\renewcommand{\thefootnote}{\fnsymbol{footnote}}
\makebox[2cm]{}\\[-1in]
\begin{flushright}
\begin{tabular}{l}
TUM/T39-96-25
\end{tabular}
\end{flushright}
\vskip0.4cm
\begin{center}
  {\Large\bf
    On the analytical approximation to the GLAP evolution at small $x$ and
    moderate $Q^2$\footnote{Work supported in
    part by BMBF}}\\ 

\vspace{2cm}

L.\ Mankiewicz\footnote{On leave of absence from N. Copernicus
Astronomical Center, Polish Academy of Science, ul. Bartycka 18,
PL--00-716 Warsaw (Poland)} A. Saalfeld and T. Weigl 

\vspace{1.5cm}

{\em Institut f\"ur Theoretische Physik, TU M\"unchen, Germany}

\vspace{1cm}

{\em \today}

\vspace{1cm}

{\bf Abstract:\\[5pt]} \parbox[t]{\textwidth}{ Comparing the numerically
  evaluated solution to the leading order GLAP equations with its analytical
  small-$x$ approximation we have found that in the domain covered by a large
  fraction of the HERA data the analytic approximation has to be augmented by
  the formally non-leading term which has been usually neglected. The corrected
  formula fits the data much better and provides a natural explanation of some
  of the deviations from the $\sigma$ scaling observed in the HERA kinematical
  range.  }

\vspace{1cm}
{\em To appear in Physics Letters B}
\end{center}
\end{titlepage}

\newpage

The high-precision data on the deep inelastic structure function $F_2^p(x,Q^2)$
coming from ZEUS \cite{ZEUS} and H1 \cite{H1} experiments installed in the HERA
accelerator have stimulated new interest in the properties of the QCD dynamics
in this kinematical domain \cite{Theo96}. The conceptually simplest analysis
typically relies on solutions of the GLAP equations which are either evaluated
numerically or approximated in the small-$x$ region by a suitable analytical
formula. In particular, Forte and Ball \cite{Forte95,Forte96} pointed out that
the HERA small-$x$ data may be interpreted in terms of the so called double
asymptotic scaling phenomenon related to the asymptotic behaviour of the GLAP
evolution first discovered by DeRujula et al.\cite{DeRuj} many years ago. On
the other hand, various groups have been able to fit the available
data using input characterized by the hard small-$x$ behaviour $x^{-(1 +
\lambda)}$,  $\lambda > 0$ \cite{varGroup}.

Notwithstanding the dispute about applicability of the double asymptotic
scaling to the HERA data, the problem of an analytical approximation to the
GLAP evolution is interesting in itself \cite{Gehr96}. A common sense requires
that any analytical formula has to follow the full solution with an accuracy at
least as good as the magnitude of the experimental error bars, a goal which may
not be so easy to achieve given the high precision of the HERA data.
Analytical approximations to the GLAP evolution are usually designed to be
valid in an extreme kinematics of very small $x$ and/or large $Q^2$.  Starting
from this observation, we have considered in the present note the question to
which extent the approximation which leads to the double scaling behaviour can
be trusted in the domain of moderate $Q^2$ of the order of $ \sim 10$ GeV$^2$
which contains a non-negligible fraction of the small-$x$ HERA data, even
assuming the most favorable case of absolutely soft initial conditions.  As we
shall show in the following, the self-consistent approximation to the GLAP
evolution in this region indeed seems to require taking into account a formally
non-leading term which is normally neglected. Alternatively, one may of course
try to avoid the problem by going to higher and higher $Q^2$, which inevitably
results in a lower statistics, especially in the small $x$ region.  On the
other hand as we show in the following, the improved formula not only provides
a much better fit to the small $x$ data, but also the appropriately corrected
data points are much more consistent with the asymptotic $\sigma$ scaling.

Our analysis is performed at leading order, but its generalization to
NLO is not particularly difficult. However, before such an extended
analysis is performed one has to be sure that there are no other sources of
corrections which can be even larger than the difference between the LO and NLO
formula \cite{Inprep}.

At leading order the solution of GLAP equations for moments
of the quark singlet $q_s(x,t)$ parton distribution
\begin{equation}
q_s(n,t) =  \int_0^1 dx\, x^n \, q_s(x,t)
\label{moments}
\end{equation}
where $t = \ln(\frac{Q^2}{\Lambda^2})$, has the well known form \cite{QCDbook}
\begin{eqnarray}
q_s(n,t) & = & \left( (1-h_2(n)) q_s(n,t_0) - h_1(n) g(n,t_0) \right)
 \exp\left[\frac{2}{\beta_0} \lambda_+(n) \zeta\right]  \nonumber \\ 
& + & \left( h_1(n) g(n,t_0) + h_2(n) q_s(n,t_0) \right)
\exp\left[\frac{2}{\beta_0} \lambda_-(n) \zeta\right] 
\label{solution}
\end{eqnarray}
where $g(n,t)$ is the corresponding moment of the gluon distribution and $\zeta
= \ln{(\frac{\alpha_s(Q_0^2)}{\alpha_s(Q^2)})}$. The coefficients $h_1(n)$ and
$h_2(n)$ as well as the eigenvalues of the anomalous dimension matrix
$\lambda_\pm$ arise from the diagonalisation of the flavour-singlet evolution
equation. To obtain the Bjorken-$x$ distribution from (\ref{solution}) one has
to take the inverse Mellin transformation i.e., evaluate the integral
\begin{equation}
x q_s(x,t) = \frac{1}{2 \pi i} \, \int_{c - i \infty}^{c + i \infty} dn x^{-n}
q_s(n,t) \, ,
\label{Mellin}
\end{equation}
where the integration contour runs to the right to all singularities of
$q_s(n,t)$. At small-$x$ one expects that the bulk of the integral comes from
the part of the contour in the vicinity of the point $n=0$, which can be
quantified e.g., by the saddle point approximation. As the flavour non-singlet
part is subleading at sufficiently small-$x$ values, in this region the
structure function $F_2(x)$ should be related to a good accuracy to the
singlet quark distribution (\ref{Mellin}) via $ F_2(x) = \frac{5}{18} \,
q_s(n,t)$ ($n_f = 4$).  In Ref.\cite{Forte95,Forte96} the HERA data were
analyzed with the help of a further analytical approximation to the integral
(\ref{Mellin}), based on the assumption that in the small-$x$ region and for
sufficiently large $Q^2$ the second term in (\ref{solution}), driven by the
eigenvalue $\lambda_-$, can be neglected with respect to the first one.  In
this limit the solution of the GLAP equation in the moment space reads
\begin{equation}
q_s(n,t) = q_+(n,t) = \left( (1-h_2(n)) q_s(n,t_0) - h_1(n) g(n,t_0) \right)
 \exp\left[\frac{2}{\beta_0} \lambda_+(n) \zeta\right] ,
\label{doubleS}
\end{equation}

The double-scaling solution discussed in Ref.\cite{Forte95,Forte96} is 
obtained by assuming soft initial conditions  $q_s(n,t_0) \sim g(n,t_0)
\sim 1/n$, expanding $\lambda_+(n)$ and $h_i(n)$ (i = 1,2) around the point 
$n=0$, retaining only the leading terms, and evaluating the resulting Mellin 
integral (\ref{Mellin}). As such simple initial conditions typically 
overestimate the actual magnitude of $x q(x,Q_0^2)$ one can introduce an
additional parameter $x_0$ which has an interpretation related to the 
constant term in the Laurent expansion of the initial conditions around their 
rightmost singularity:
\begin{equation}
q(n,t_0) = A(Q_0^2)\,\left(\frac{1}{n} + \ln{x_0}\right) + \dots =
 A(Q_0^2)\, \frac{1}{n} \exp{(n\ln{x_0})} + \dots\, .
\label{init1}
\end{equation} 
Assuming that $x q(x,Q_0^2) \sim (1-x)^\beta$ as $x \to 1$, with $\beta \sim 3$
one expects $x_0 \sim e^{-2} \sim 0.1$. 

The reasoning which leads to the approximation (\ref{doubleS}) is based on the
series of assumptions which have to be fulfilled by the data under scrutiny.
Hence, to be able to justify the small-$x$ and large $Q^2$ approximations we
have selected for the further analysis 83 points in the domain $x \le 0.01$,
$Q^2 \ge 5$ GeV$^2$ from the recently published H1 data \cite{H1}.  Although
for some data in this domain the value of $Q^2$ is large, relatively many
points have much lower $Q^2$, of the order of $10$ GeV$^2$. On the other hand,
if the perturbative QCD evolution is to be valid, the starting point of the
$Q^2$ evolution should not be much smaller than, say, 1 GeV$^2$.  Then, as the
first consistency check we have plotted the distribution of the saddle point
positions $n_0 = (\frac{12}{\beta_0} \frac{\zeta}{\xi})^{1/2}$,
$\xi=\ln{(x_0/x)}$, for $x_0 = 1$ and $x_0 = 0.1$, see full and shadowed
histograms on Figure 1 respectively. It is interesting to see that especially
in the second case saddle points are not very close to zero, so that one can
doubt whether the usual small $n$ approximation to the integral (\ref{Mellin})
is applicable at all. To quantify this problem, in the next step we have
compared the results of the full solution \cite{Wei96} to the evolution
equation (\ref{solution}) with the double-scaling approximation
(\ref{doubleS}) at $Q^2 = 10$ GeV$^2$, assuming the soft initial conditions for
quark and gluon distribution functions $q_s(n,t_0) = \frac{A}{n}$, $g(n,t_0) =
\frac{B}{n}$ at $Q_0^2 = 1$ GeV$^2$, taking for simplicity $A=B$. The results
are shown on Figure 2. The data points correspond to $Q^2 = 8.5$ GeV$^2$, and
the normalization of initial conditions has been simply adjusted to the data by
hand. It turns out that even for such simple initial conditions the
double-scaling approximation, represented by the dashed line, approaches the
full solution, depicted by the solid line, at values of $x$ much smaller than
those experimentally accessible at present. Moreover, in the region covered by
the data the difference between the full solution and the double-scaling
approximation is significantly larger than the error bars of the data points,
making the applicability of the latter to the analysis of the data at least
questionable. The situation does not change if we assume more realistic initial
conditions, say, $x q_s(x) \approx x g(x,t_0) = (1-x)^3$, see Figure 3 where
the appropriate $x_0$ parameter has been taken into account in the
double-scaling approximation. Our simplifying choice $A=B$ cannot be
responsible for the discrepancy - a consistent analytical approximation has to
be able to accommodate various initial conditions without compromising the
accuracy. 

Comparing the situation for various values of $Q^2$ we have found that the
situation improves considerably when $Q^2$ increases, or in other words in the
above discussion we have considered, for the sake of clarity of the argument, a
really extreme example. By the same token one can suspect that for a reasonable
description of relatively low $Q^2$ data the term driven by the eigenvalue
$\lambda_-$ should not be neglected. We have tested this hypothesis by
comparing again the full solution with the approximation in which both the
first and the second terms in (\ref{solution}) are expanded in $n$ to the
leading order, i.e.
\begin{eqnarray}
q_s(n,t) & = & \left(\, 0.198 \, n\, q_s(n,t_0) + 0.444\, n \,g(n,t_0)\,
 \right)
 \exp\left[ \frac{2}{\beta_0} (6/n-5.648)\, \zeta \right]   \nonumber \\ 
& + & q_s(n,t_0) \exp\left[-\,\frac{2}{\beta_0}\, 1.185 \, \zeta \right] \, . 
\end{eqnarray}
All real coefficients above result from expansion of corresponding coefficients
in (\ref{solution}) in $n$ and keeping only the first terms.  After
transformation into $x$-space the final expression for the structure function
$F_2(x)$ to be fitted to the H1-data in the following reads:
\begin{eqnarray}
 F_2(x,t )& = & \frac{5}{18} \, \left(\, 0.198 \, A + 0.444 \, B\, \right)
 \exp\left[- \frac{2}{\beta_0} 5.648 \, \zeta \right]  
 \left(\frac{\gamma^2 \,\, \zeta}{\xi} \right)^{\frac12}
 I_1\left(2\, \gamma  \sqrt{\zeta \xi} \right) \nonumber \\ 
 & + & \frac{5}{18} \, A \, \exp\left[-\,\frac{2}{\beta_0}\, 1.185 \, 
 \zeta \right] \, , 
\label{solution1}
\end{eqnarray}
where $\gamma = \sqrt{12/\beta_0}$. As it can be seen from Figure 4, the
formula (\ref{solution1}) considerably improves the quality of the
approximation, which in our opinion makes its applicability much better
justified.

Motivated by these considerations we have compared fits to the small-$x$ data
set based on formulae (\ref{doubleS}) and (\ref{solution1}) respectively.
For that purpose we have used the minimization package MINUIT from the
CERN library \cite{MINU}.
The results are summarized in Tables 1 and 2. We have kept $n_f=4$, used
$\Lambda_{QCD} = 250$ MeV and a fixed input scale $Q_0^2 = $ 1GeV$^2$ such that
pQCD should be still meaningful. The improved approximation
(\ref{solution1}) results in $\chi^2$ which is smaller approximately by a
factor 3. We note that if we had not restricted $Q_0^2$ to a fixed value of
$Q_0^2 = 1$ GeV$^2$, the fit would drive $Q_0^2$ to much lower values of about
$0.5$ GeV$^2$, simultaneously pushing $x_0$ to values larger than 1, which is
certainly disturbing as far as the interpretation of $x_0$ is concerned. On the
other hand $\chi^2$ hardly changes by such a shift, and hence we are sure that
the values presented in Table 1 represent a sensible set of fit parameters.  In
addition, if the data points in the $Q^2$ bins around $5$ GeV$^2$ and $6.5$
GeV$^2$ are removed from the data set, the fit improves once more considerably,
see corresponding entry in Table 2.

\begin{table}[h]
{\hfill
\begin{tabular}{|c|c|c|c|} \hline
 data range & $0.198 \,A + 0.444 \, B$ & $x_0$ &  $\chi^2/d.o.f.$ \\ \hline
  $Q^2 \ge 5$ GeV$^2$, $x \le 0.01$   & 1.60 $\pm$ 0.06 
  & 0.14 $\pm$  0.01  & 1.57  \\ \hline
  $Q^2 \ge 8.5$ GeV$^2$, $x \le 0.01$  &  1.41 $\pm$ 0.06 
  & 0.19 $\pm$ 0.02  & 1.35\\ \hline
\end{tabular}
\hfill}
\caption{LO fit with double-scaling solution}
\label{LOdoubleS}
\end{table}

\begin{table}[h]
{\hfill
\begin{tabular}{|c|c|c|c|c|} \hline
 data range  & $A$ & $B$ & $x_0$ &  $\chi^2/d.o.f.$ \\ \hline
  $Q^2 \ge 5$ GeV$^2$, $x \le 0.01$   & 0.83 $\pm$ 0.09 
  & 2.47 $\pm$ 0.18   & 0.15 $\pm$ 0.02  & 0.49  \\ \hline
  $Q^2 \ge 8.5$ GeV$^2$, $x \le 0.01$ &   0.80 $\pm$ 0.10  
  & 2.30 $\pm$ 0.18 & 0.18 $\pm$ 0.03  & 0.35 \\ \hline
\end{tabular}
\hfill}
\caption{LO fit with improved solution}
\label{LOsolution1}
\end{table}

What is the influence of the correction taken into account in (\ref{solution1})
on the double asymptotic scaling considered in \cite{Forte95,Forte96}? To see
this we have compared the LO double asymptotic scaling formula which results
from (\ref{doubleS}) with the data in the domain $x \le x_0$, $Q^2 \ge 8.5$
GeV$^2$ before and after the correction due to the extra term in
(\ref{solution1}) is taken into account. The so called $\sigma$-scaling
predicts \cite{Forte95} that after rescaling the $F_2$-data by the factor
\begin{equation}
 R_F = N_F \, \rho \, \sqrt{\sigma} \exp(- 2 \, \gamma\, \sigma\, + 
       \delta_+ \, \sigma / \rho)
\label{dscale}
\end{equation}
where $\delta_+ = (11 + \frac{2}{27} n_f)/\beta_0$, one should observe constant
behaviour at large $\rho$, where $\rho = \sqrt{\xi/\zeta}$ , $\sigma =
\sqrt{\xi \, \zeta}$, $\gamma = \sqrt{12/\beta_0}$) and $N_F$ is an arbitrary
normalization constant.  The results are presented on Figure 5. The upper plot
corresponds to the original double-scaling formula (\ref{dscale}). For the
lower plot our fit result for the non-scaling contribution arising from the
formally non-leading term in (\ref{solution1}) has been subtracted from each
data point before the rescaling. For $\rho \ge 1.5$, points on the lower plot
clearly have much smaller spread around the horizontal line. The $\rho$
scaling predicts that on the plot $\ln(R_{F'} \, F_2)$ 
versus $\sigma$ the data points, with the rescaling factor given by
\begin{equation}
R_{F^\prime} = N_{F'} \rho \sqrt{\sigma} \exp(\delta_+ \, \sigma / \rho)
\label{dscale1} 
\end{equation}
should lie on the straight line with the slope equal to $2 \gamma$. To see the
effect of (\ref{solution1}) on the $\rho$ scaling hypothesis we have displayed
on Figure 6 the scaling behaviour before (upper plot) and after (lower plot)
the non-leading correction has been subtracted from the data points. Although
both plots clearly differ in the domain of $\sigma \le 1$, the scaling
behaviour is hardly influenced for $\sigma \ge 1.5$. The dashed line represents
the expected behaviour according to the first, scaling term in
(\ref{solution1}) i.e.,
\begin{equation}
R_{F^\prime} F_2 = N \sqrt{\sigma} I_1(2 \gamma \sigma)
\end{equation}
with the constant $N$ adjusted to the normalization of the large $\sigma$
data points.

For large $\sigma$ the data plotted in Fig.6 should follow a straight line
with slope $2 \gamma$ \cite{Ball94}.
Restricting our data range to $Q^2 \ge 8.5$ GeV$^2$, $x \le 0.01$ we have
determined, using MINUIT as well,
the slope before and after the non-leading correction has been
subtracted from the data points. Our results, summarized in Table 3, clearly
show that with the non-leading correction taken into 
account we get better agreement with the value expected from 
the theory, $2 \gamma_{th} = 2.4$

\begin{table}[h]
{\hfill
\begin{tabular}{|c|c|c|} \hline
        & data range & $2 \gamma$  \\ \hline
  double scaling solution & $\sigma \ge 1.0$   & 2.15  $\pm$ 0.04 \\ \hline
  improved solution &   $\sigma \ge 1.0$  &  2.47 $\pm$ 0.05 \\ \hline
\end{tabular}
\hfill}
\caption{Slope $2 \gamma$ obtained from a fit to $F_2$-data}
\label{LOslopes}
\end{table}

To summarize, starting from the observation that the experimental errors set
the magnitude for the minimal sensible accuracy of an analytical approximation
to the GLAP evolution, we have found that the description in the region of $Q^2
\sim 10$ GeV$^2$ improves considerably after a formally non-leading, and thus
usually neglected term is included. The improved formula provides a better
description of the available small $x$ HERA data and it improves the observed
$\sigma$ scaling.

\vskip 1cm

{\bf Acknowledgments}\\
This work was supported in part by BMBF, KBN grant 2~P03B~065~10 and
German-Polish exchange program X081.91.

\vfill\eject

\clearpage

%%%%%%%%%%%%%%%%%% REFERENCES %%%%%%%%%%%%%%%%%%%%%%%%%%%%%%

\clearpage

%%%%%%%%%%%%%%%%%%%%%% Figures %%%%%%%%%%%%%%%%%%%%%%%%%%%%%%%%%%

%\epsfbox{figure1.ps}
\begin{minipage}{14cm}
\bild{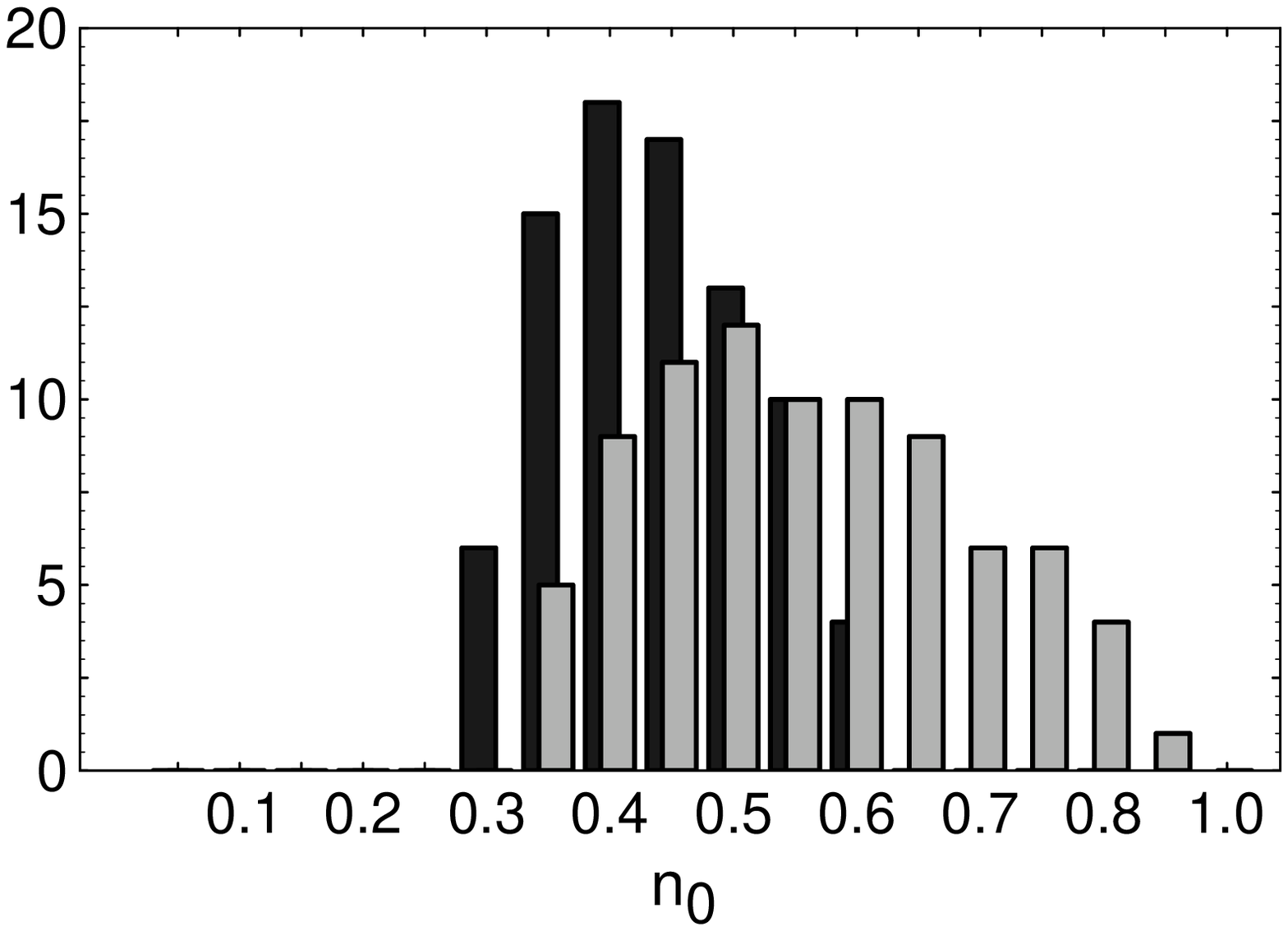}{14}
\end{minipage}
\begin{description}
  
\item[Fig.~1] Distribution of saddle points $n_0 = (\frac{12}{\beta_0}
  \frac{\zeta}{\xi})^{1/2}$ over the analyzed HERA data points \cite{H1}. The
  shadowed and filled histograms correspond to $x_0=0.1$ and $x_0=1$,
  respectively.

\end{description}

\clearpage

\epsfbox{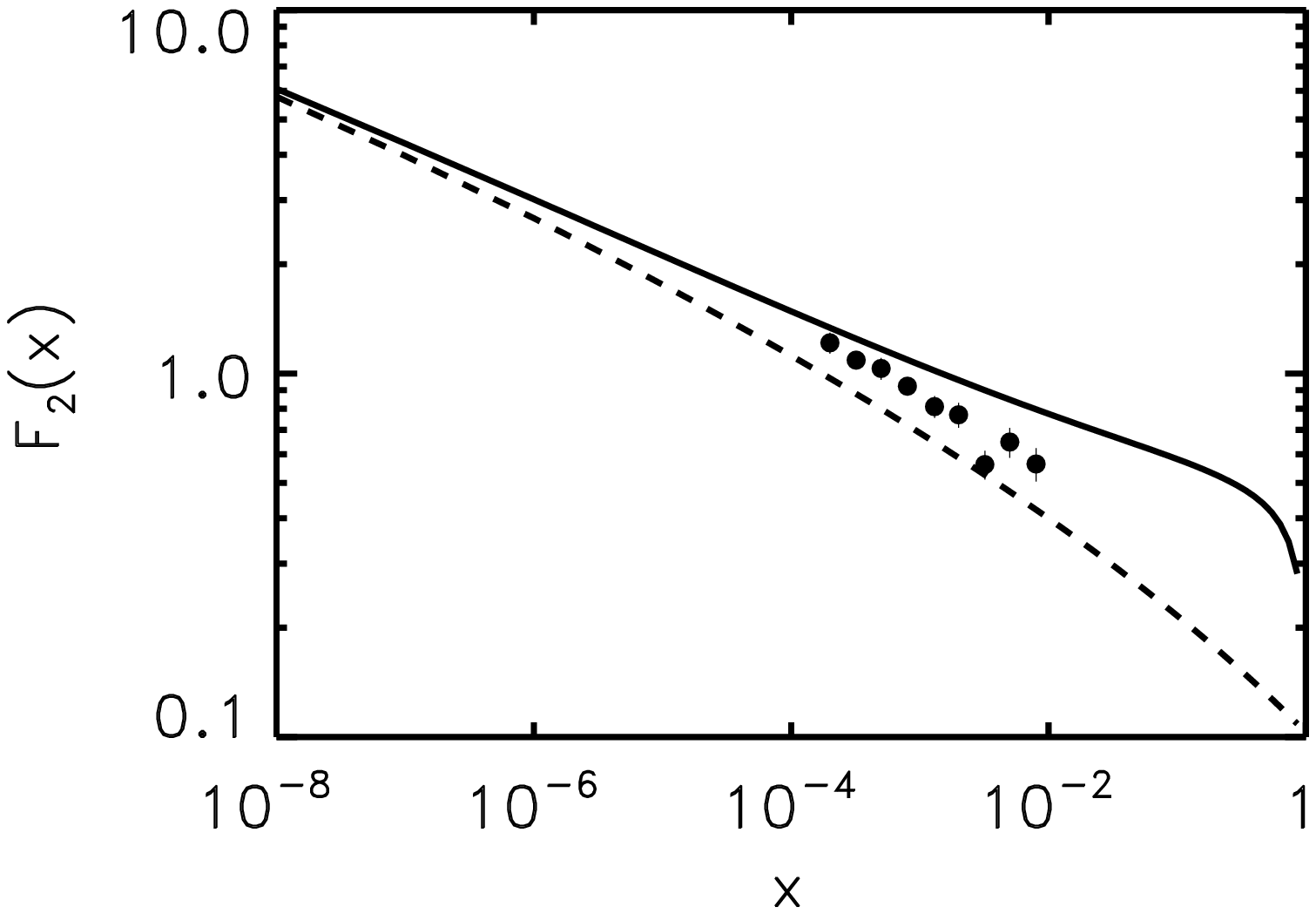}

\begin{description}
  
\item[Fig.~2] Comparison of the full evolution (solid line), equation 
  (\ref{solution}) with
  the approximate double-scaling solution (\ref{doubleS}) for the data around
  $Q^2 = 10$ GeV$^2$ \cite{H1}, and the soft initial conditions $q(n,t_0) =
  g(n,t_0) = 0.45/n$ at $Q_0^2 = 1$ GeV$^2$.

\end{description}

\clearpage

\epsfbox{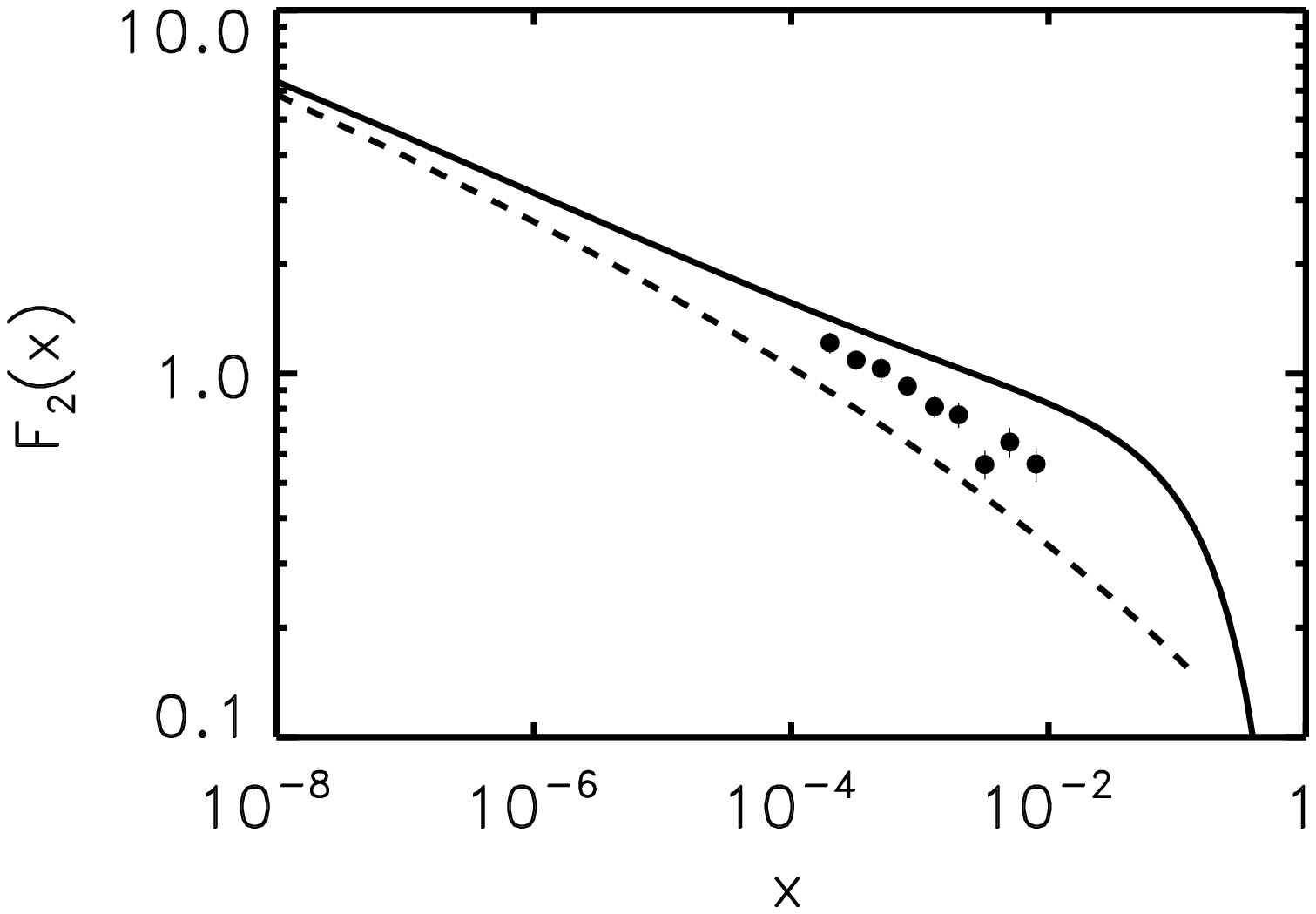}
\begin{description}
  
\item[Fig.~3] Comparison of the full evolution (solid line), equation 
  (\ref{solution}) with
  the approximate double-scaling solution (\ref{doubleS}) for the data around
  $Q^2 = 10$ GeV$^2$ \cite{H1}, and the soft initial conditions $x q(x,t_0) = x
  g(x,t_0) = 0.6 \, \, (1-x)^3$ at $Q_0^2 = 1$ GeV$^2$.

\end{description}

\epsfbox{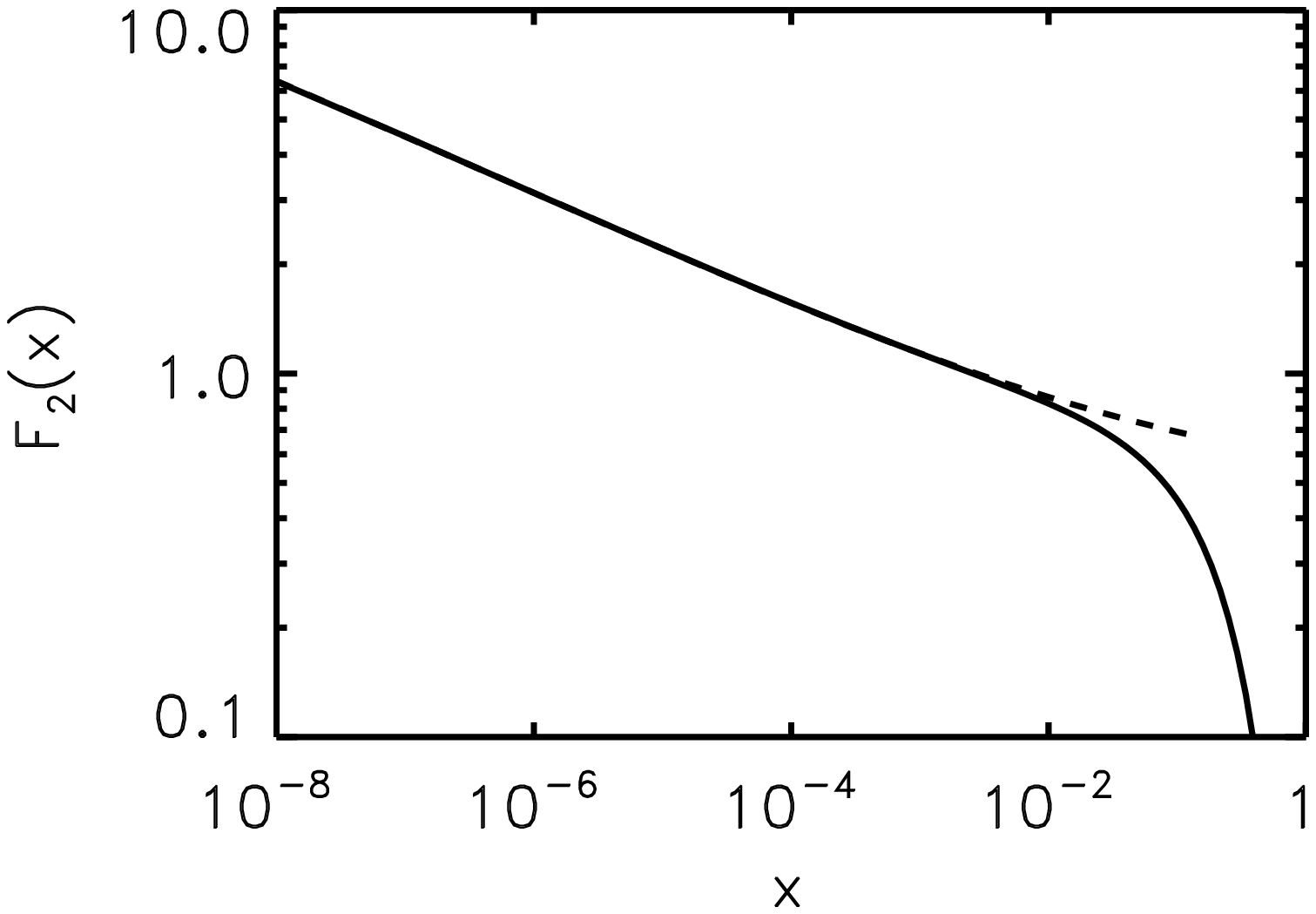}
\begin{description}
  
\item[Fig.~4] Comparison of the full evolution (solid line), equation 
  (\ref{solution}) with
  the approximation which includes besides the double-scaling solution also
  the leading contribution from the term driven by the non-leading eigenvalue
  $\lambda_-(n)$, equation (\ref{solution1}), for the soft initial
  conditions $x q(x,t_0) = x g(x,t_0) = 0.6 \, \,(1-x)^3$ at 
  $Q_0^2 = 1$ GeV$^2$.

\end{description}

%\epsfbox{figure5.ps}

\begin{minipage}{13cm}
\bild{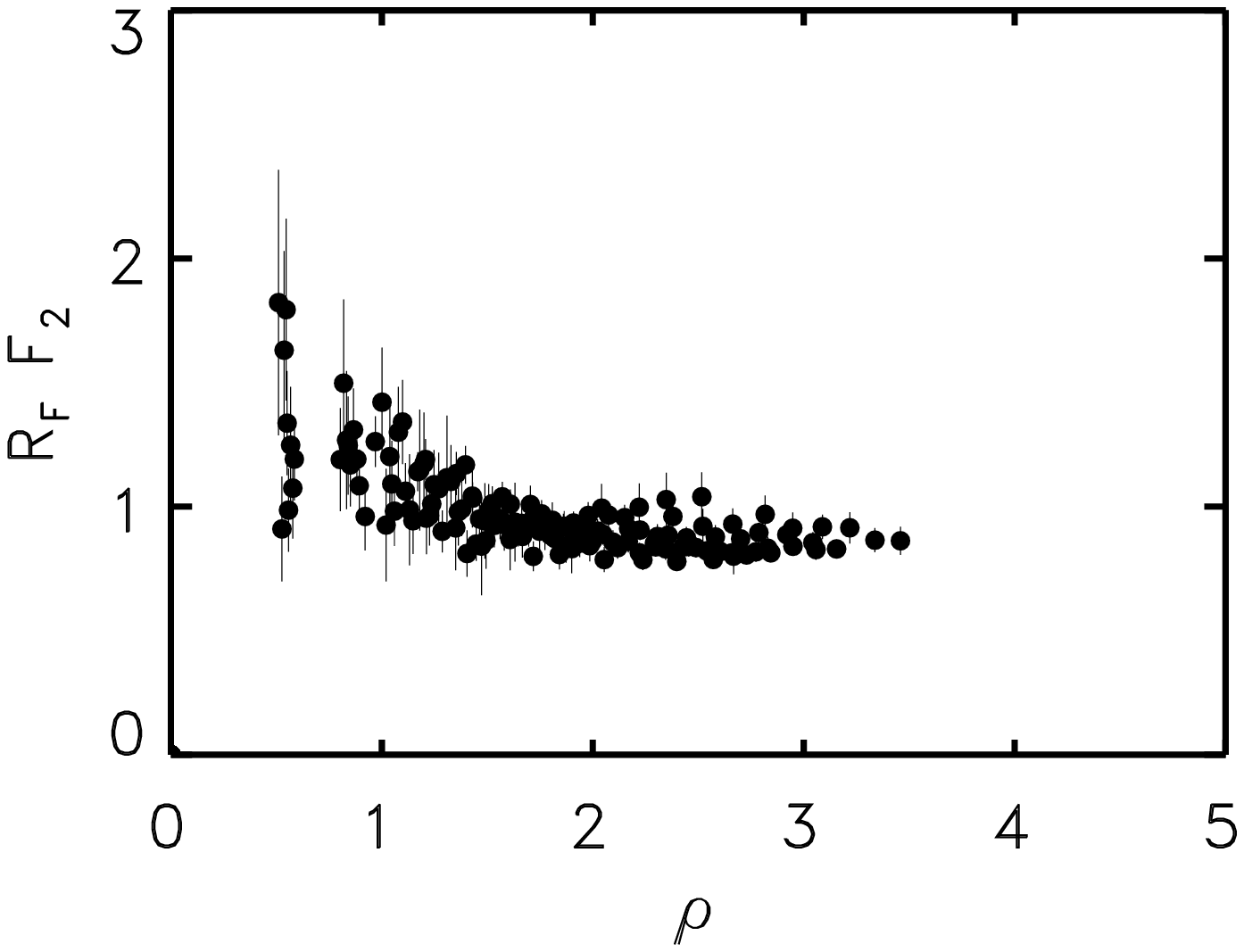}{13}
\bild{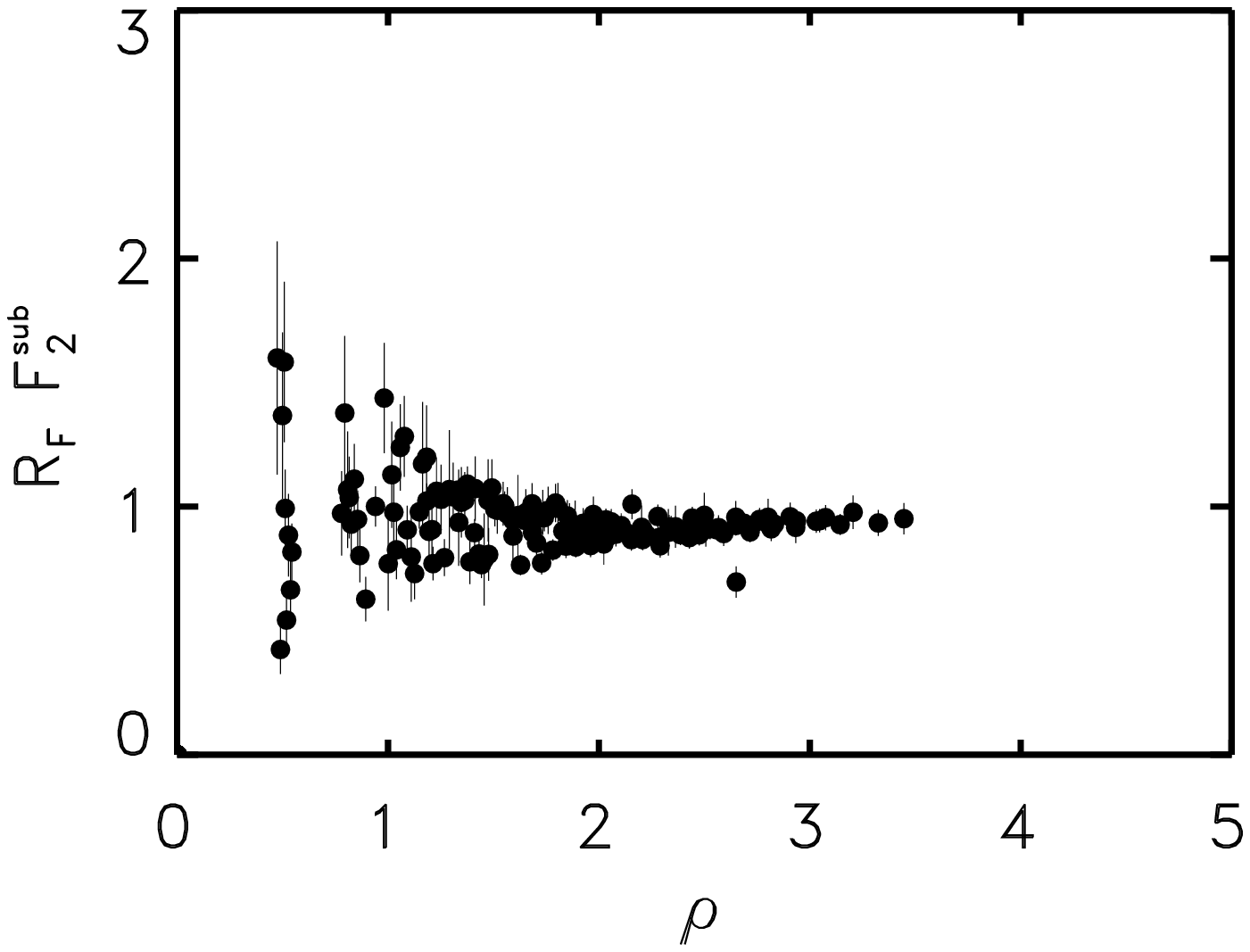}{13}
\end{minipage}

\begin{description}
 
\item[Fig.~5] The $\sigma$ scaling formula from Ref.\cite{Forte95,Forte96}
  versus the small $x$ HERA data \cite{H1}. For the lower plot the formally
  non-leading contribution taken into account in (\ref{solution1}) has been
  subtracted from each data point before rescaling of the data \cite{H1}
  according to (\ref{dscale}).

\end{description}

%\epsfbox{figure6.ps}

\begin{minipage}{13cm}
\bild{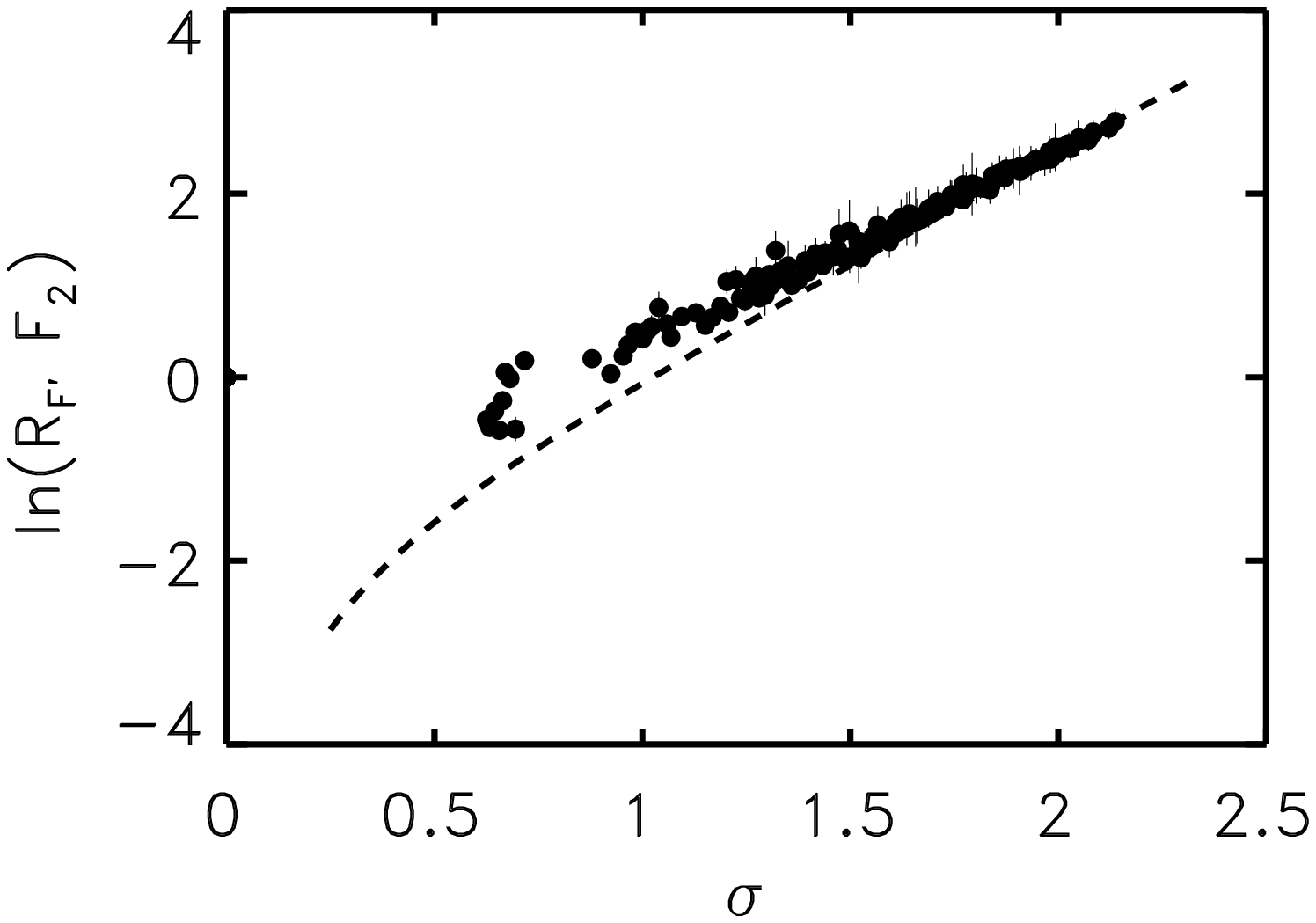}{13}
\bild{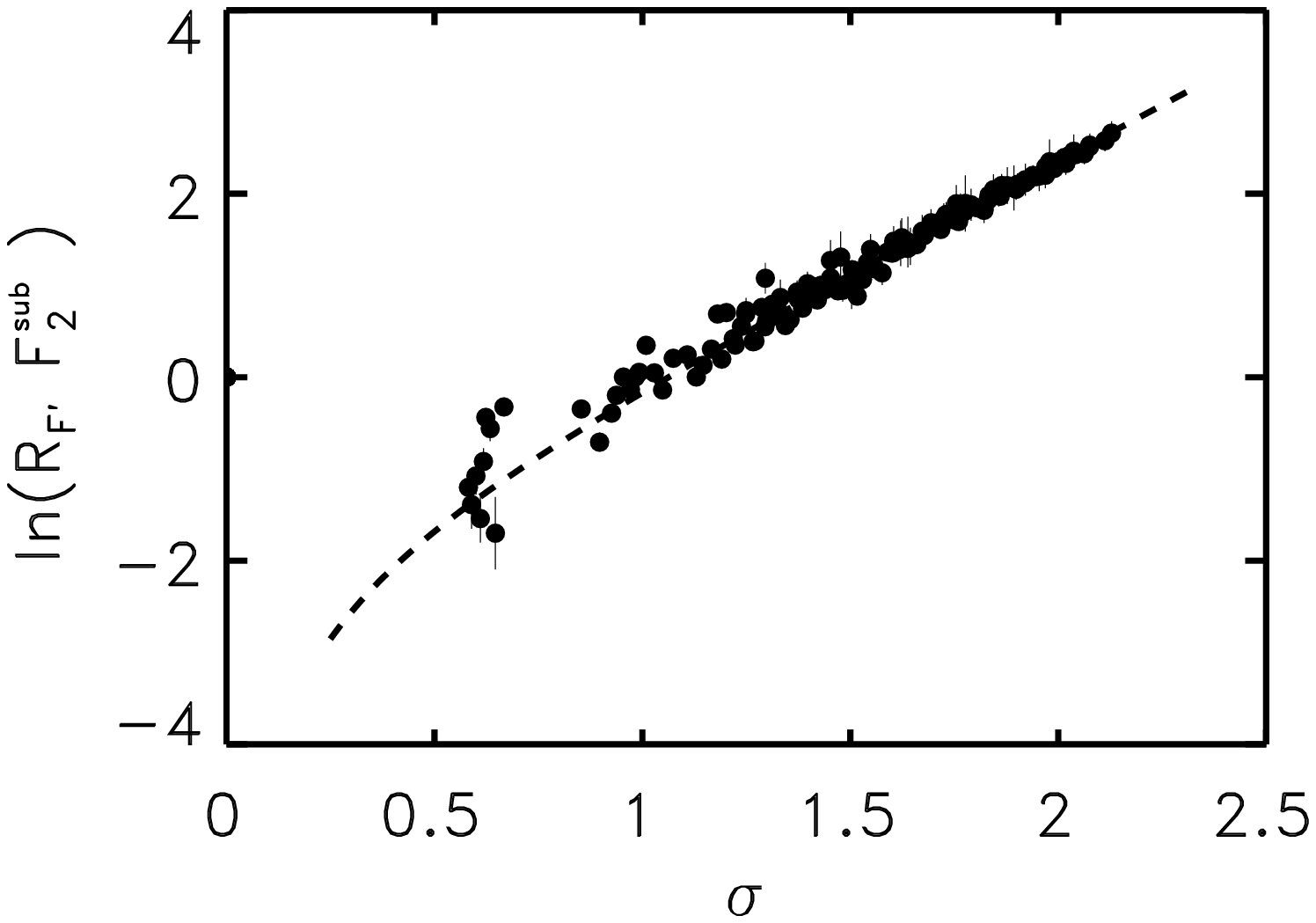}{13}
\end{minipage}

\begin{description}
 
\item[Fig.~6] The $\rho$ scaling formula from Ref.\cite{Forte95,Forte96} versus
  the small $x$ HERA data \cite{H1}. For the lower plot the formally
  non-leading contribution taken into account in (\ref{solution1}) has been
  subtracted from each data point before rescaling of the data \cite{H1}
  according to (\ref{dscale1}). The dashed line represents the expected
  behaviour according to the first, scaling term in (\ref{solution1}).

\end{description}

\end{document}